\documentclass[a4paper]{article}
\usepackage{amsmath,amssymb}
\newcommand{\ssy}[5]{#1, {#2} {\bf #3}, #5 (19#4)}
\newcommand{\rnum}[1]{\mathrm{I\!R^#1}}
\newcommand{\kart}[4]{\begin{figure}[#1]\begin{center}\leavevmode%
\epsfysize=#2\epsffile{#3.ps}\end{center}\caption{#4}\end{figure}}
\newenvironment{bukv}{\refstepcounter{equation}
\setcounter{table}{\value{equation}}\setcounter{equation}{0}
}
{\setcounter{equation}{\value{table}}}%	
\title{A singularity-free WEC-respecting time machine}
\author{S. V. Krasnikov\thanks{Email: \it redish@pulkovo.spb.su}\\
The Central\\
Astronomical Observatory at Pulkovo, St Petersburg,\\
196140, Russia}
\date{}
\begin{document}
\input{epsf}
\maketitle
\begin{abstract}
A time machine (TM) is constructed whose creating in contrast to all
TMs known so far requires neither singularities, nor violation of
the weak energy condition (WEC). The spacetime exterior to the TM closely
resembles the Friedmann universe.\\[\medskipamount]PACS numbers:
04.20.Dw, 04.20.Gz
\end{abstract}  
\section{Introduction}
This paper concerns an aspect of the long-standing question: How to
create a time machine (or why is it 
impossible)? We begin with the following
\paragraph{Definition.} Let $N$ be an inextendible acausal spacetime.
We call $L_N\subset N$ a \emph{time machine} (created in the universe
$M$) if
\begin{enumerate}
	\item  $L_N$ comprises the causality violating set $V$:
	$$L_N\supset V\equiv\{\mathcal P\mid\mathcal P\in N,
	\:J^+(\mathcal P)\cap J^-(\mathcal P)\neq\mathcal P\}$$
	\item $N\setminus J^+(L_N)$ is isometric to $M\setminus J^+(L_M)$,
	 where $M$ is some inextendible causal spacetime and 
	$\overline{L_M}\subset M$ is compact. 
\end{enumerate}\par\noindent
It is meant here that depending on whether  we decide to make
a time machine or not the world and our laborotory must be described
by $N$ and $L_N$, or by $M$ and $L_M$, respectively. 
We require the compactness of $L_M$ to differentiate TMs,
which supposedly can be built by some advanced civilization, and
causality violations of a cosmological nature such as the G\"odel
universe or the  Gott ``time machine" \cite{Cos}. 
\par
A few important facts are known about time machines. Among them:
\begin{enumerate}
	\item Time machines with compact $V$ (\emph{compactly generated TMs,
	CTMs}) evolving from a noncompact partial Cauchy
	surface must violate WEC \label{Haw} \cite{Conj},
	\item Creation \label{Tip} of a CTM leads to singularity formation
	unless  some energy condition (slightly stronger than WEC) is 
	violated\footnote{
		 Moreover, it is not clear whether singularities can be 
		avoided at all, even at the sacrifice of WEC. 
	Absence of singularities has not been proven, 
	as far as I know, for any of TMs considered so far.}
	\cite{Tip}.
\end{enumerate}
Though strictly speaking these facts do not \emph{prove} that CTMs are
impossible altogether, they, at least, can be used as starting
point for the search for a mechanism protecting causality from CTMs
\cite{Conj}. \par
CTMs, however, only constitute a specific class of TMs. Noncompactly
generated TMs (NTMs) seem to be every bit as interesting as CTMs.
Sometimes (see e.~g.~\cite{Conj}) they are barred from consideration
on the basis that some unpredictible information could enter
noncompact $V$ from a singularity or from infinity. This is so
indeed, but compactness does not eliminate this trouble (in fact,
compactness does not even eliminate singularities as a possible
source of unpredictible information, as in the Misner space). The
creation of a time machine is inherently connected with a loss of
predictability 
(cf.~\cite{Par}).  One inevitably risks meeting something unexpected 
(i.~e.~not fixed by the data on the initial surface) as soon as one
intersects a Cauchy horizon (by the very definition of the horizon).
So, CTMs and  NTMs are not too different in \emph{this}
regard.\par
It therefore seems important to find out whether there are any similar
obstacles to creating  NTMs.  The example of the Deutsch-Politzer
TM \cite{Deu} showed that item \ref{Haw} is not true in the case of
NTMs.  This TM, however, possesses singularities (and though mild,
they are of such a nature that one cannot ``smooth" them out \cite{Sin}).
Thus the following questions remained unanswered:
\begin{enumerate}
	\item Do time machines without singularities exist?
	\item Are the weak energy condition and the absence of
	singularities mutually exclusive for (noncompactly generated) 
 	time machines?
\end{enumerate}
Our aim in this paper is to give the answers to both these questions
(positive to the first and negative to the second). We make no
attempt to discuss possible consequences of these answers. In
particular, being interested only in the very existence of the
desired TM we consider the fact that it may be created
(see the Definition) in the spacetime resembling the Friedmann universe,
only as a pleasant surprise.
\section{Construction of the time machine}%                        SECT I
To construct a singularity-free TM it would be natural to start from
the Deutsch-Politzer TM and to look 
for an appropriate conformal transformation of its metric which would
move the dangerous points to infinity. However, it is not easy in the
four-dimensional case to find
a transformation (if it exists) yielding both WEC fulfilment and
$b$- (or BA-) completeness. So, we shall use somewhat
different means \cite{Ya}. First, by a conformal transformation we
make a part of the \emph{two-dimensional}
Deutsch-Politzer spacetime locally complete (the spacetime $Q_N$
below), then compose $L_N$ obeying WEC from $Q_N$ and some $S$
(chosen so that it does not spoil the completeness), and finally embed
the resulting TM in an appropriate $N$.\par
%
%			CURVED D-P TM
%
\subsection{Curved Deutsch-Politzer time machine $Q_N$}
Let $Q_{f}$  be a square $q_m$
\begin{equation} %                                               Lim
	q_m\equiv\{\chi,\tau\,\bigl|\; m>|\chi|,|\tau|\}
	\label{Lim}
\end{equation}
endowed with the following metric:
\begin{equation}%                                             Gq
	ds^2=f^{-2}(\tau,\chi)(-d\tau^2+d\chi^2)	
	\label{Gq}
\end{equation}
Here $f$ is a smooth bounded function defined on $\rnum{2}\supset
q_m$ such that 
\begin{equation}%                                                F1
	f(\tau,\chi)=0\quad \Leftrightarrow \quad 
	\{ \tau=\pm h,\; \chi=\pm h \}
	\label{F1}
\end{equation}
with $0<h<m/2$. The four points $f^{-1}(0)$ bound two segments
$$
	l_\pm\equiv\{\chi,\tau\,\bigl|\; \tau=\pm h,\:|\chi|<h\}
$$
and we require  that
\begin{equation} %                                               F2
	{f(\tau,\chi)|\,}_{U^+}=f(\tau-2h,\chi),
	\label{F2}
\end{equation}
where $U^+$ is some neighborhood of $l^+$.

Now (as is done with the Minkowski plane in the case of the ``usual''
Deutsch-Politzer spacetime \cite{Deu}) remove the points $f^{-1}(0)$
from $q_m$, make 
cuts along $l_+$ and $l_-$, and glue the upper bank of each cut with
the lower bank of the other (see Fig.~1).
%\begin{figure}[!h]
%\centering
%\unitlength=0,24pt
%\begin{picture}(640,640)
%\put(-20,1000){\special{em:graph fig1stm.pcx}}
%\end{picture}
%\caption
\kart{!h}{30ex}{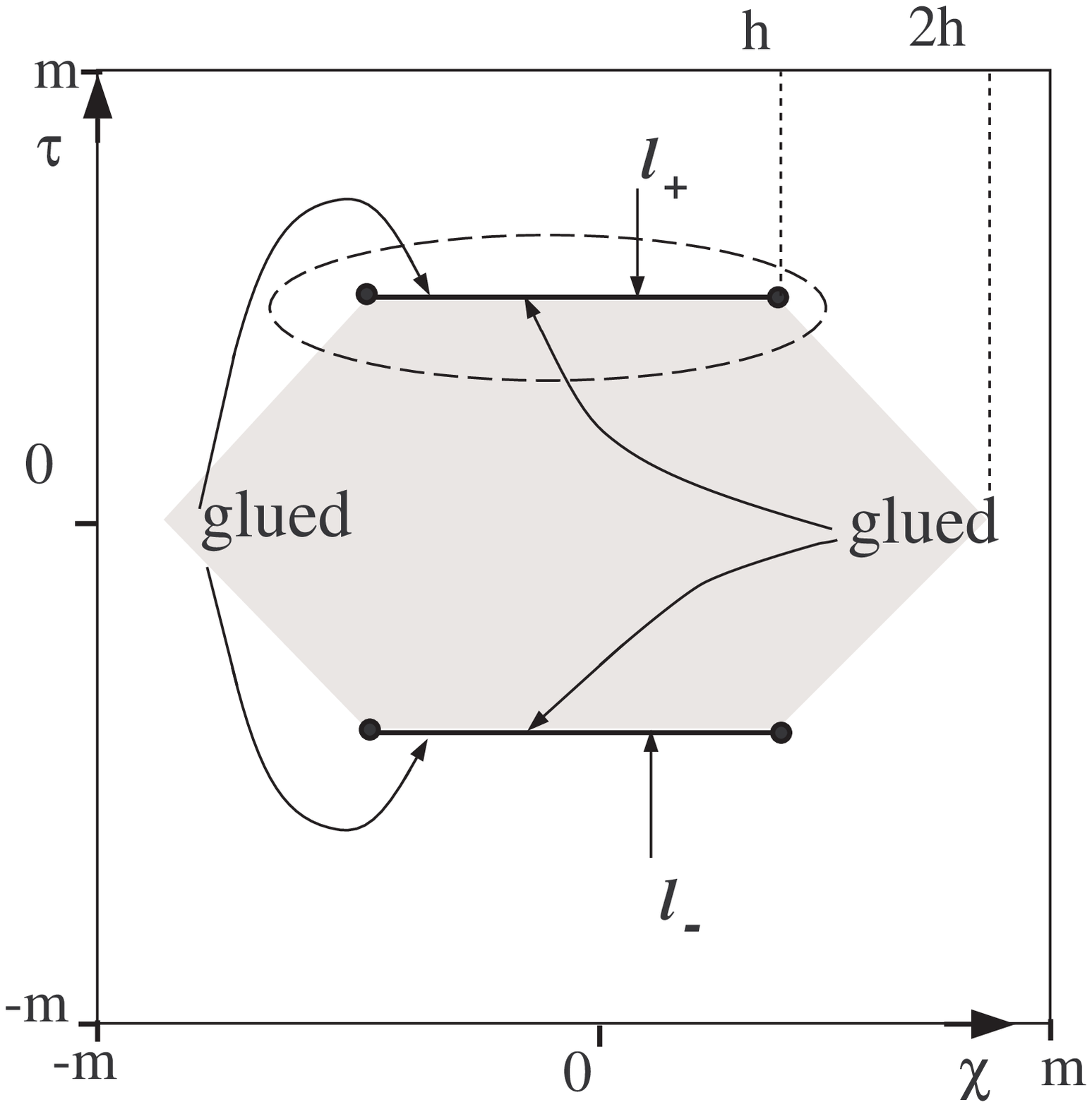}%
{Curved Deutsch-Politzer time machine. The shaded region is
the causality violating set. The dashed line bounds $U^+$.}
%\end{figure}
 The resulting spacetime
$Q_N$ (``curved Deutsch-Politzer TM") is not, of course, diffeomorphic to
$Q_{f}$. 
Nevertheless, for simplicity of notation we shall continue to use
the ``old coordinates'' $\tau,\chi$ for its points.
%
%			TIME MACHINE
%
\subsection{The time machine $L_N$}
Let $S$ be a two-sphere with the standard metric:
\begin{equation}%                                            Sp
	\label{Sp}
	ds^2=R^2(d\theta^2+\sin^2 \theta d\varphi^2)
\end{equation}
where $R$ is a constant satisfying (to bring about  WEC
fulfillment, see below)
\begin{equation}%                                            WEC
	\label{WEC}         
R^{-2}\geqslant\max_{\overline q_m}\bigl(f(f,_{\chi\chi}-f,_{\tau\tau})
	+f,_\tau^2 - f,_\chi^2\bigr)
\end{equation}
Then
\begin{equation}%                                            LN
	L_N\equiv Q_N\times S
	\label{LN}
\end{equation}
is just the desired time machine.  
\subsection{Exemplary spacetimes $N,\,M$}
To find an appropriate $N$
require in addition to (\ref{F1},\ref{F2},\ref{WEC})
\begin{equation}%                                                F3
	f(\mathcal P)=const\equiv f_0\quad
	\mbox{ \rm when }\mathcal P \notin q_m
	\label{F3}
\end{equation}
and denote by $\widetilde Q$ the spacetime obtained by
replacing $q_m\to\rnum{2}$ in the definition of $Q_N$. It follows
from what is proven 
in the next section that 
the spacetime $\widetilde Q\times S$ is
inextendible and could thus be taken as $N$ (with, for example,
$L_M\equiv Q_{f_0}\times S$).
 We would like, however, to
construct another, more ``realistic'' $N$.

Consider a manifold $\rnum{1}\times \mathrm S^3$ with the metric 
\begin{equation}%                                              Fr
	ds^2=a^2[-d\tau^2+d\chi^2+\rho^2
	(d\theta^2+\sin^2 \theta d\varphi^2)]
	\label{Fr}
\end{equation}
Here $\tau$ is a coordinate on $\rnum{1}$ and $\theta,
\:\varphi,\:\chi\,(-\pi/2\leqslant\chi\leqslant\pi/2$) are polar
coordinates on S$^3$. Impose the following conditions on $a,\,\rho$
(it suffices to choose $m<\pi/2,\:f_0<R^{-1}\cos m$ for their
feasibility): 
\begin{bukv}
\begin{align}%                                              AI,RI
	\text{ on }q_m\qquad&&
	 a&=1/f,&\rho&=fR \label{Qn}\\
	\text{ exterior to }q_m\qquad&&
	 a&=\hat a(\tau),&\rho&=\hat \rho(\chi),
	\label{A1}
\end{align}
\end{bukv}
where $\hat \rho,\,\hat a$ are \emph{convex} positive functions and for some
$n\in(m,\pi/2)$ holds $\hat \rho|_{|\chi|>n}=\cos \chi$.
\par
It is easy to see that the region $|\tau|,\,|\chi|<m$ of this manifold
 is $Q_{f}\times S$.
%\begin{figure}[!h]
%\centering
%\unitlength=0,24pt
%\begin{picture}(640,680)
%\put(-20,1000){\special{em:graph fig2stm.pcx}}
%\end{picture}
%\caption
\kart{!h}{30ex}{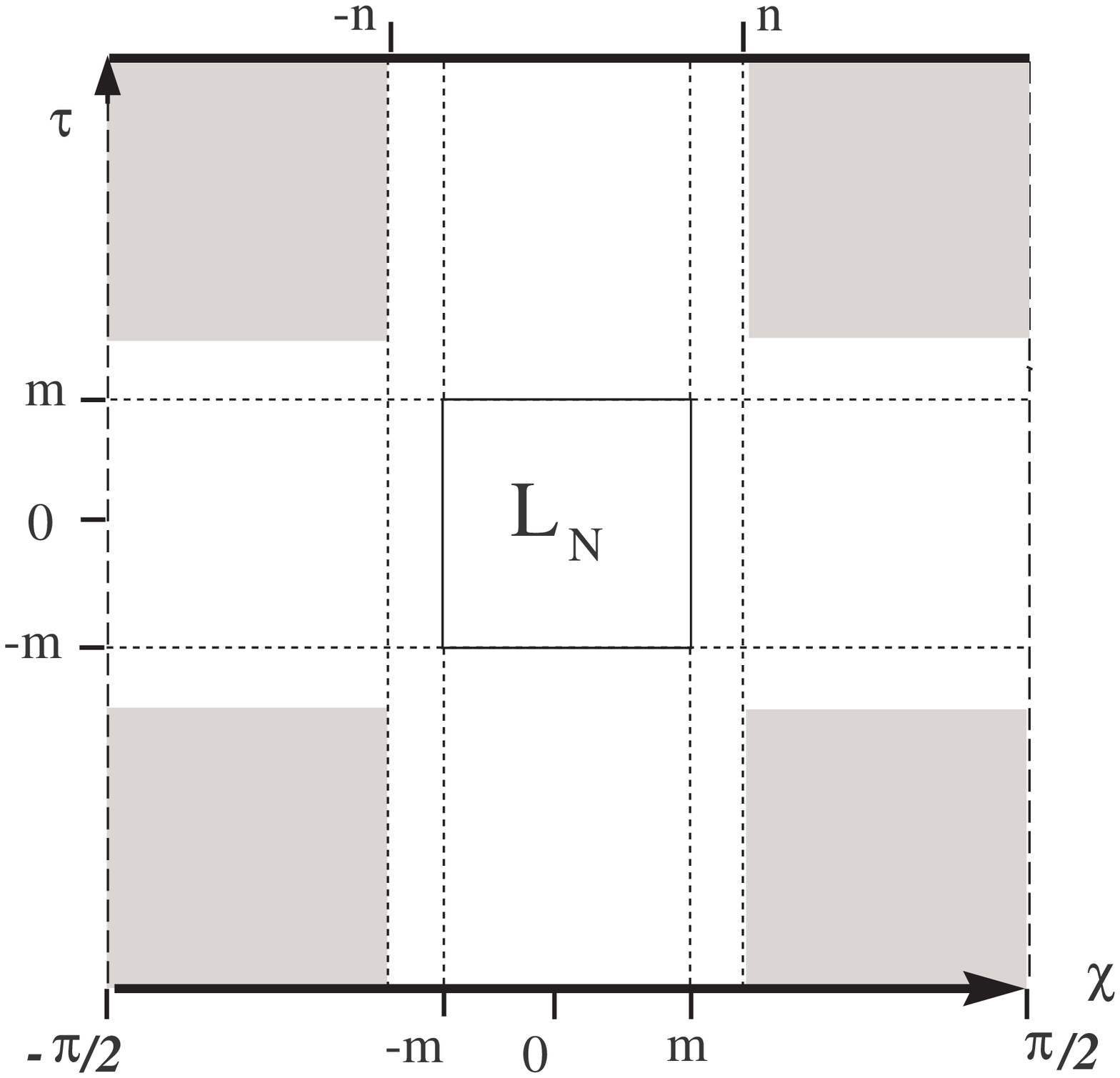}%
{``Almost Friedmann'' time machine. Shaded regions are
parts of the Friedmann universe. The thick horizontal lines depict
cosmological singularities.}
%\end{figure}
So we can repeat the manipulations with cuts and obtain a TM with the metric 
(\ref{Fr}) on $N\setminus \overline{L_N}$ (see Fig.~2), that is the
TM is created in a spacetime with the metric of
  the Friedmann universe outside some spherical
layer and some time interval. 

\section{Proofs} %                                     SECTION 2

\subsection{Weak energy condition}
The metric of the time machine ${L_N}$ due to (\ref{Qn}) is given by 
(\ref{Gq},\ref{Sp}) and the condition (\ref{WEC}) guarantees that the
weak (and even the dominant) energy conditions hold there (see
\cite{Ya} for details).
\par In the outer space $N\setminus
{L_N}$ the metric is given by (\ref{Fr}). Introducing the
quantities 
	$$
	\Phi \equiv \Lambda \equiv \ln a,
	\quad r \equiv a \rho
	$$
we bring it to the form (14.49) of ref.~\cite{MTW}. The fact that by
(\ref{A1})  $a$
and $\rho$ are positive and convex  gives us:
	\begin{equation}
	\ddot\Phi \leqslant 0, \quad \rho''/\rho \leqslant 0, \quad
	1 - {\rho'}^2 \geqslant 0,
	\label{Pos}
	\end{equation}
(in the last inequality we have also used that $\rho'(\pm\pi/2)=\mp 1$.)
Hence (see \cite{MTW} for notation)
	\begin{eqnarray}
	&&E=\bar E= a^{-2}\ddot\Phi \leqslant 0, \quad H=0,\nonumber\\
	&&F=r^{-2}(1 - {\rho'}^2) +  
	a^{-2}{\dot\Phi}^2 \geqslant 0,\nonumber\\
	&&\bar F = a^{-2} ({\dot\Phi}^2 - \rho''/\rho) \geqslant 0\nonumber
	\end{eqnarray}
So (see (14.52) of ref.~\cite{MTW}), WEC holds in this region too.

\subsection{Completeness}
The results of \cite{Ya} prove that there are no
``BA-singularities'' in $L_N$, that is any timelike inextendible (in
$N$) curve $\gamma\subset L_N$ with bounded acceleration has infinite
proper length. There is a popular idea, however, that only
$b$-complete regions may be accepted as singularity-free. So, the
remainder of the article is devoted to the proof of the fact that
$L_N$ (not the whole $N$, where cosmological singularities $\hat a=0$
present) has no ``$b$-singularities.'' We shall use the following new
notation:
\begin{gather*}
		x^1\equiv \tau, \quad x^2\equiv \chi, \quad
		x^3\equiv\theta, \quad x^4\equiv\varphi,\\
		\alpha\equiv \chi+\tau,\quad
		\beta\equiv \chi-\tau,\quad\dot{}\equiv d/ds.
\end{gather*}
Let $\gamma(s)=x^i(s),\:i=1,\dots 4$ be a $C^1$ curve in $L_N$. 
It defines two other curves (its projections onto $Q_N$ and $S$):
\begin{eqnarray}%                                              Gam
	Q_N\supset&\gamma_Q(s)\equiv x^k(s),&\quad k=1,2\\
	S\supset&\gamma_S(s)\equiv x^j(s),&\quad j=3,4
\end{eqnarray}%                                    
Lying in $L_N$ and $Q_N$ the curves $\gamma$ and $\gamma_Q$ can be
considered at the same time as 
lying in $N$ and $\widetilde Q$, respectively. We shall call such curves 
inextendible if they are inextendible in those ``larger" spacetimes.
Let \{\textbf e$_{(i)}(s)$\} be an orthonormal basis in the point $\gamma(s)$,
obtained from \{\textbf e$_{(i)}(0)$\} by parallel propagating along
$\gamma$ and \textbf e$_{(\alpha)}\equiv
	\mathbf e_{(1)}+\mathbf e_{(2)},\:
	\mathbf e_{(\beta)}\equiv
	\mathbf e_{(1)}-\mathbf e_{(2)}$.
 Choosing \textbf e$_{(i)}(0)\sim\partial_{x^i}$
 and solving
the equations $\nabla_{\dot\gamma}\mathbf e_{(i)}=0$ one immediately
finds 
\begin{equation}%                                              Ind
	\nabla_{\dot\gamma_Q}\mathbf e_{(k)}=0,\quad
	\nabla_{\dot\gamma_S}\mathbf e_{(j)}=0
	\label{Ind}
\end{equation}
and
\begin{equation}%                                             Ek
	e^i_{(\mu)}(s)=
	e^i_{(\mu)}(0)\exp\{2\int_0^s\phi,_\mu\dot\mu\,ds'\}
 	\quad (\mu\equiv\alpha,\beta)
	\label{Ek}
\end{equation}
where $\phi\equiv\ln f$.

The ``affine length'' $\mu_N$ of $\gamma$ is by definition \cite{Sch1}
\begin{equation}%                                            Mu
	\mu_N[\gamma]\equiv\int_0^1\Bigl(\sum_i
	\langle\dot\gamma,\textbf e_{(i)}\rangle^2\Bigr)^{1/2}ds
	\label{Mu}
\end{equation}
We define affine lengths $\mu_Q[\gamma_Q]$ and $\mu_S[\gamma_S]$ by
changing $i\,$ in (\ref{Mu}) to $k$ and $j$, respectively. Due to
(\ref{Ind}) these definitions are consistent. Obviously,
\begin{equation}%                                            Otz
	\mu_N[\gamma]\geqslant1/2\,(\mu_Q[\gamma_Q]+\mu_S[\gamma_S])
	\label{Otz}
\end{equation}
\par\medskip
\textbf{Proposition.} If $\gamma\subset L_N$ is inextendible %       PROP
 then $\mu_N[\gamma]=\infty$.

\smallskip\noindent
If $\gamma$ is inextendible, than either
$\gamma_S$ or $\gamma_Q$ (or both) are inextendible, too. But $S$ is
obviously $b$-complete. So the Proposition follows from (\ref{Otz})
coupled with the following

\par\medskip\textbf{Lemma.} If $\gamma_Q$ is inextendible, %           LEMMA
 then $\mu_Q[\gamma_Q]=\infty$.

\smallskip\noindent
Let us introduce a function $\Phi$ on $\gamma_Q$:
$$
	\Phi\equiv\int_0^s
	(\phi,_{\alpha}\dot\alpha-\phi,_{\beta}\dot\beta)\,ds'
$$
and let us split $\gamma_Q$ on segments
$\gamma_n\equiv\gamma[s_n,s_{n+1})$ so that the sign of $\Phi$ does
not change on $\gamma_n$:
\begin{equation}%                                         Spli
	\gamma_Q=\bigcup_n\gamma_n,\quad
	\gamma_n:\Phi(s_n)=0,\:
	\Phi(s_n<s<s_{n+1})\leqslant0\:(\mbox{or}\,\geqslant0).
	\label{Spli}
\end{equation}
Denote the contribution  of a segment $\gamma_n$ in $\mu_Q[\gamma_Q]$
 by $\mu_n$. Since
$$
	\left(
	\langle\dot\gamma,\textbf e_{(1)}\rangle^2+
	\langle\dot\gamma,\textbf e_{(2)}\rangle^2\right)^{1/2}
	\geqslant 1/2\left(
	|\langle\dot\gamma,\textbf e_{(\alpha)}\rangle|+
	|\langle\dot\gamma,\textbf e_{(\beta)}\rangle|
	\right)
$$
we can write for  $\mu_n$ (cf. (\ref{Ek})):
\begin{equation}%                                           Otz2
	\mu_n\geqslant C_1\int\limits^{s_{n+1}}_{s_n}f^{-2}\Bigl(
	|\dot\alpha|\exp\{2\int_0^s\phi,_\beta\dot\beta\,ds'\} +
	|\dot\beta|\exp\{2\int_0^s\phi,_\alpha\dot\alpha\,ds'\}
					\Bigr)\,ds
	\label{Otz2}
\end{equation}
Here and subsequently we denote by $C_p,\;p=1,\dots$ some irrelevant
positive constants  factored out from the integrand.
Using
\begin{equation}%                                           Tozh
	\bigl[f(s)\bigr]^{-2}=\bigl[f(0)\bigr]^{-2}\exp\{-2\int_0^s
	(\phi,_\alpha\dot\alpha + \phi,_\beta\dot\beta)\,ds'\}
	\label{Tozh}
\end{equation}
we can rewrite (\ref{Otz2}) as
\begin{equation}%                                           Otz3
	\mu_n\geqslant C_2\int\limits^{s_{n+1}}_{s_n}\Bigl(
	|\dot\alpha|\exp\{-2\int_0^s\phi,_\alpha\dot\alpha\,ds'\} +
	|\dot\beta|\exp\{-2\int_0^s\phi,_\beta\dot\beta\,ds'\} 
					\Bigr)\,ds
	\label{Otz3}
\end{equation}
For definiteness let $\Phi\leqslant0$ on $\gamma_n$. 
Then the first exponent in (\ref{Otz3}) is greater than the second
and we can replace it by their geometric mean, that is [see
(\ref{Tozh})] by $f(0)/f(s)$. So,
\begin{multline}%
	\mu_n\geqslant C_3\int^{s_{n+1}}_{s_n}|\dot\alpha/ f|\,ds\geqslant
	C_4\int^{s_{n+1}}_{s_n}|\phi,_\alpha\dot\alpha|\,ds\\
	\geqslant-C_5\int^{s_{n+1}}_{s_n}(\phi,_\alpha\dot\alpha+
	\phi,_\beta\dot\beta)\,ds\geqslant 
	C_5\bigl(\phi(s_n)-\phi(s_{n+1})\bigr)
	\label{Otzf}
\end{multline}
(The third inequality follows again from $\Phi\leqslant0$.)
Clearly, (\ref{Otzf}) also holds  for those $\gamma_n$ where
$\Phi\geqslant0$ and hence summing over $n$ gives:
\begin{equation}%                                              Poln
	\mu_Q[\gamma_Q]\geqslant C_5\bigl(\phi(0)-\phi(s)\bigr),
	\quad\forall s\in[0,1)
	\label{Poln}
\end{equation}
There are no closed null geodesics in $\widetilde Q$. So, it is ``locally
complete" \cite{Sch2}. That is any inextendible $\gamma_Q$ either has
$\mu_Q[\gamma_Q] =\infty$,
or leaves any compact subset of $\widetilde Q$. But in the latter case
(recall that $\gamma_Q\in Q_N$) $\phi(s)$ is unbounded below, which
due to (\ref{Poln}) again gives 
$$
\mu_Q[\gamma_Q]=\infty.
$$
\hfill$\square$

\section*{Acknowledgement}

This work is partially supported by the RFFI grant 96-02-19528.

\end{document}